# About Network Structures and Systems Complexity

Olexandr Polishchuk

Laboratory of Modeling and Optimization of Complex Systems, Pidstryhach Institute for Applied Problems of Mechanics and Mathematics National Academy of Sciences of Ukraine

Lviv, Ukraine

od_polishchuk@ukr.net

*Abstract*— This paper provides the analysis for structural and functional approaches of complex network systems research. In order to study the behavior of these systems the flow adjacency matrices were introduced, and local and global dynamic characteristics of system elements were defined. The notion of the flow core of network system was introduced and its advantages over the k-core of complex network were analyzed. The methods were proposed for identifying real structure of network system and reducing the dimensionality of its model.

## I. INTRODUCTION

To study any real system, whether natural or artificial, we have to form full and comprehensive representation of this system. Usually it is reached through observations, experimental and theoretical investigations and displaying the system as the models of different types (informational, structural, functional, mathematical etc.) [1]. While modeling large-scale complex systems we face the number of problems. First of all, it is a qualitative complexity that derives from incomplete understanding of laws according to which many systems operate [2]. Qualitative complexity of the system is also based on the fact that processes taking place within the system depended on numerous factors. Indeed, in real world, the same event can have multiple reasons and same reasons under different conditions may cause different events. These peculiarities usually result in models being nonlinear and their application quite difficult. Another problem consists in quantitative complexity or dimensionality of both real systems and their models [3]. The network systems (NSs) belong to the category of so-called condensed environments that can be studied only as a whole [4]. That is, the dimensionality of NS model should be proportional to the dimensionality of the network system itself. If the system has $10^6$ to $10^{12}$ elements and operation of each element is described by only one equation, the system of equations of the same dimensionality will be obtained. Creating and solving systems of nonlinear equations with the dimensionality of 1 billion is a super-difficult problem. Quantitative complexity sometimes acquires the features of qualitative complexity. This happens when it is impossible to comprehend all the processes occurring in the system due to the excessive amount of data that describe those processes. This phenomenon was called information overload or analysis paralysis. While reducing the dimensionality of models of those systems, for example by the methods of decomposition, we often lose their adequacy [5]. During the study of many systems researchers face the problem that consists in continuous and rapid change of system structure and operation. For such NSs, the adequacy of the model whose creation or improvement would require certain amount of time may also be lost. The purpose of this article is to develop methods for flow analysis of complex network systems operation and ways to reduce dimensionality of their models while simultaneously monitoring the adequasy of such models.

## II. NETWORK STRUCTURES AND SYSTEMS

Complexity of network structures and systems as well as of their models in general may be represented by different concepts. Network complexity is determined, in particular, by the presence of a large number of nodes and edges between them [6]. Networks with relatively small number of elements (hundreds, thousands, and even dozens of thousands) are usually not considered complex. Research team can count dozens of workers, human musculoskeletal system consists of 206 bones and 640 muscles, Ukrainian railway transportation network includes about 1600 nodes, i. e. railway stations etc [7, 8]. But these relatively small structures generate unquestionably complex systems. In other words, the complexity of the network structure is quantitative, and the complexity of the system is qualitative. While trying to embrace functional complexity we often have to neglect the structural complexity. Among the examples of this situations are the attempts to solve problems associated with controllability and observability of NSs. At the present stage, such problems are being solved for the simplest linear models of network systems with the number of nodes up to 100 [9]. Such structures are hard to be called complex. At the same time, problems associated with controllability, observability and synchronization of large-scale systems are rather complex functional, not structural problems.

When talking about systems research, two approaches may be distinguished: structural and functional. In modern NSs studies, the structural approach prevails, which is implemented in so-called theory of complex networks (TCN) [10, 11]. The subject of TCN studies is the creation of universal network structures models, determination of statistical features that characterize their behavior and forecasting networks behavior in case their structural properties change. Sometimes the term *complex network* (CN) is used to denote both structure and system [12, 13], though these are fundamentally different concepts. The laws according to which the systems operate are usually much more complicated than the features of system structure, and methods of structural studies often do not allow us to solve NS functional problems. The subject of functional approach to systems research is the study of different system types and classes, basic system behavior principles and patterns, as well as processes of goal setting, operation, development and interaction with external environment [14]. Within the scope of functional approach, system structure is analyzed in conjunction with functions implemented by components of this structure and system in general, but the function takes precedence over structure. This does not downplay the significance of structural approach of studies, as long as poor operation of many real systems is driven by the disadvantages of their structure [7, 15]. Moreover, some systems have existed and do exist now that can be studied through exploring only their structure [16].

The network structure is completely determined by its adjacency matrix $\mathbf{A} = \{a_{ij}\}_{i,j=1}^{N}$, where $N$ is the number of CN nodes [11]. For the most studied binary networks, the value of $a_{ij}$ is equal to 1, if there is a connection between the nodes, and is equal to 0, if such connection is absent. Using the matrix $\mathbf{A}$ are defined the local and global characteristics of CN and studied its properties. In general, the theory of binary networks is completely abstracted from the functional features of the NS. Weighted networks are an attempt to "tied" the functional characteristics of the system to the elements of structure [17]. Indeed, in each particular case, the weight of CNs edges is a reflection of certain functionality of the corresponding system [18].

Network, as a structure, is considered to be dynamic if the composition of its nodes and edges changes over time. The system is a dynamic formation, even if its structure remains unchanged. The system forms its structure in the process of development. The structure is being developed and improved from the needs of the system and not vice versa. What prompts the structure to develop, modify, or degrade? Movement of flows is one of the defining features of real NS. In some cases, providing the movement of flows is the main goal of creation and operation of such systems (transport and telecommunication systems, resource supply systems, trade and information networks, etc.), in others – the main process that provides their vital activity (blood and lymph flows, neuronal impulses in the human body). Stopping of flows movement leads to the termination of the NS existence. In the life cycle of arbitrary real system we can distinguish at least three main stages: 1) growth, which is accompanied by the deployment of structure and organizing of flows movement; 2) stable functioning or "maturity", which is characterized by the coordinated motion of flows and practically unchanged NS structure; 3) "aging", which is characterized by processes that are inverse to those that occurred at the growth stage of NS. The flows analysis allows us to determine at which stage of the life cycle is the system. Indeed, if the number of nodes and edges of NS in (through) which the flows have to be directed is larger than the available number of nodes and edges, then the system is in the growth stage. Network growth is finished when the relationship between the number of required and existing NS elements becomes roughly equal and the system is in the stable operation stage. If the number of required nodes and connections becomes smaller than the available number of elements of the structure, then the system has moved to the aging stage. The model of scale-free networks development, proposed by A.-L. Barabasi and R. Albert [19], with its concepts of growth and preferential attachment sufficiently adequately reflects the processes of new territories development, implementation of new technologies, expansion of infrastructure, transport, trade, infocommunication networks, etc. However, the abstraction from the causes of structure development generated by the goal of system existence, leads to that other stages of the NS life cycle are ignored. Even at the growth stage the accounting of flows movement into network allow us to supplement the Barabasi-Albert model and more accurately determine the probability of preferential attachment of a new node or the choice of node, which will be the first attached to the NS structure from the set of applicant nodes. This clarification has a concrete applied value, since establishing a new connection is sometimes a long and costly process [16].

### III. NETWORK SYSTEMS AND FLOWS

We describe the process of system functioning on the basis of flows motion analysis by the network and introduce the following adjacency matrices of NS:

1) the matrix of the density of flows which are moving by the network edges at the current moment of time $t$:

$$\mathbf{\rho}(t,x) = \{\rho_{ij}(t,x)\}_{i,j=1}^{N}, x \in (n_i, n_j),$$

where $(n_i, n_j)$ is the edge connected network nodes $n_i$ and $n_j$, $i, j = \overline{1, N}$, $t > 0$;

2) the matrix of volumes of flows that are moving by the network edges at time $t$:

$$\mathbf{v}(t) = \{v_{ij}(t)\}_{i,j=1}^{N}, v_{ij}(t) = \int_{(n_i, n_j)} \rho_{ij}(t, x) dl,$$

3) the integral flow adjacency matrix (IFAM) of volumes of flows passed through the network edges for the period [0, $T$] to the current moment $t$:

$$\mathbf{V}(t) = \{V_{ij}(t)\}_{i,j=1}^{N}, V_{ij}(t) = \tilde{V}_{ij}(t) / \max_{m,l=\overline{1,N}} \tilde{V}_{ml}(t), \ \tilde{V}_{ij}(t) = \int_{t-T}^{t} v_{ij}(\tau) d\tau, t \geq T > 0,$$

4) the matrix of loading of network edges at time $t$:

$$\mathbf{u}(t) = \{u_{ij}(t)\}_{i,j=1}^{N}, u_{ij}(t) = v_{ij}(t) / v_{ij}^{\max},$$

where $v_{ij}^{\max}$ is bandwidth of the edge connected the network nodes $n_i$ and $n_j$, $i, j = \overline{1, N}$,

5) the integral matrix of NS loading for period [0, $T$] to the moment $t$:

$$\mathbf{U}(t) = \{U_{ij}(t)\}_{i,j=1}^{N}, U_{ij}(t) = \int_{t-T}^{t} u_{ij}(\tau) d\tau / T, t \geq T > 0.$$

The introduced above flow adjacency matrices in aggregate give a sufficiently clear quantitative picture of the system's operation process, allow us to analyze the features and predict the behavior of this process, to evaluate its effectiveness and prevent existing or potential threats [20, 21]. The matrix $\mathbf{\rho}(t,x)$ can be useful for the current analysis of network system's operation. In the case of transport systems, Fig. 1a shows a fairly even distribution of vehicles in a path connected the nodes $n_i$ and $n_j$, and does not pose potential threats for flows movement. Fig. 1b shows that there are obstacles on the way (points *A* and *B*), which lead to the accumulation of vehicles and may cause traffic jams. The main feature of existence on the edge $(n_i, n_j)$ the nodes that are not included in the network structure is the removing of existing or adding of new volumes of flows. If the density function behaves jump-like on the edge (Fig. 1c, points *C* and *D*), this allows us to suspect the existence of such nodes on this edge, and the behaviour of its first derivative – to identify their location (Fig. 1d).

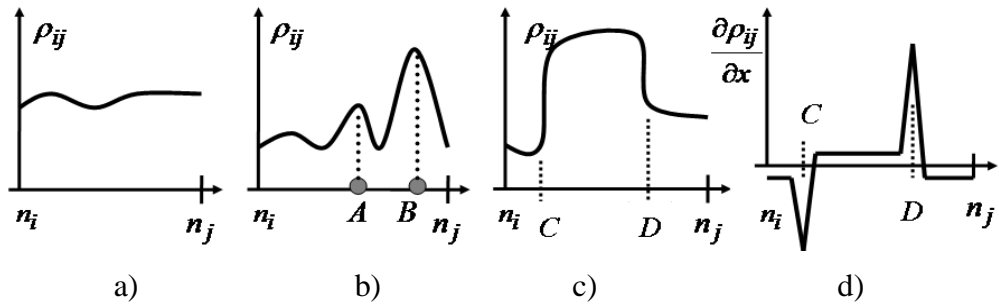

a) b) c) d)

Fig. 1. The function of flow density on the edge $(n_i, n_j)$, $i, j = \overline{1, N}$.

The matrices $\mathbf{v}(t)$ and $\mathbf{V}(t)$ enable to track the current and integral volumes of flows that pass through the network edges. They are especially important in predicting and / or planning the NS operation and allow us to timely respond to deploying threatening processes in the system. The matrices $\mathbf{u}(t)$ and $\mathbf{U}(t)$ enable to analyze the current and integral activity or passivity of separate system components, as well as the level of their critical loading, which can lead to crashes in the NS operation. These matrices allow us to timely increase the bandwidth of network elements, build new ones or search the alternative paths of flows movement, etc.

During investigation of system and forming its model we are interested in a clear identification of the NS structure. The network elements that are not involved in the system operation will be called fictitious. Examples of the existence of numerous fictitious nodes and edges can be found in many real systems, including social networks and the Internet [22]. The World Wide Web has a deep and dark web, pages of which are not indexed by any search engines [23]. Elements that are involved in the operation of particular system, but not included in its structure, will be called hidden. The identification of hidden nodes and connections plays no less important role in constructing the NS model than the search of fictitious elements. Obviously, the removal of fictitious elements contributes to overcoming the complexity problem by reducing the dimensionality of system model, and the inclusion of hidden nodes and connections – to better understanding of processes that occur in it. The flow adjacency matrices of the NS enable to identify the fictitious elements in the source network (Fig. 2a and 2b) and exclude them from the system structure (Fig. 2c). These matrices also allow to carry out the search and inclusion of hidden nodes and connections in the system structure (Fig. 2d) [16].

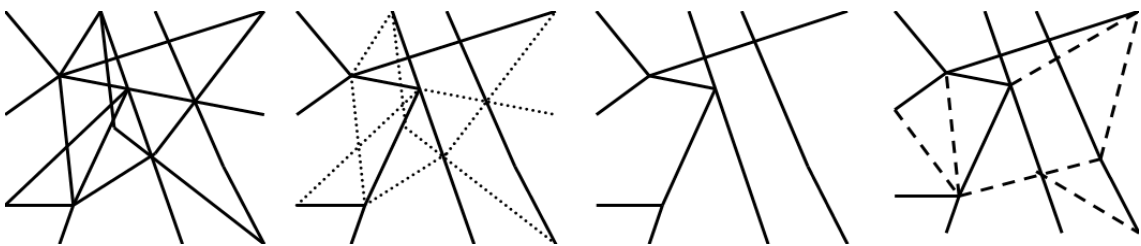

Fig. 2. Identification of fictitious and hidden CN elements (....... – fictitious edges, ------ – hidden edges).

Different ways are to determine both the local and global importance of the network node in TCN: the degree of node, betweenness centrality, average proximity to this node of all other CN nodes, etc. However, the importance of a node in the structure is often not the same as the functional significance of node in the system.

We determine the total volume of flows that arrived at node $n_i$ from node $n_j$ by edge $(n_j, n_i)$ for period [0, T] by the ratio

$$w_{ji}^{in}(t) = \int_{t-T}^{t} \rho_{ji}(\tau, x_i, y_{ji}(x_i)) d\tau, t \geq T > 0,$$

where $y = y_{ji}(x)$, $x \in [x_j, x_i]$, is the equation of a curve that connects network nodes $n_j$ and $n_i$. Denote by $L_i^{in} = \{l_i^1, ..., l_i^k\}$ the set of node numbers connected by edges with node $n_i$ from which the flows are directed to node $n_i$. Then parameter

$$W_i^{in}(t) = \sum_{j \in L_i^{in}} w_{ji}^{in}(t)$$

defines the input flow degree of node $n_i$ at the current time $t$. Parameter

$$\varphi_{ji}^{in}(t) = \frac{w_{ji}^{in}(t)}{d_i^{in}}$$

where $d_i^{in}$ is the input degree of node $n_i$ determines the importance for this node the edge $(n_j, n_i)$.

The output flow degree of node $n_i$ is determined by the ratio

$$W_i^{out}(t) = \sum_{j \in L_i^{out}} w_{ij}^{out}(t),$$

where $L_i^{out} = \{l_i^1, ..., l_i^n\}$ is the set of node numbers connected by edges with node $n_i$ to which the flows are directed from the node $n_i$ and

$$w_{ij}^{out}(t) = \int_{t-T}^{t} \rho_{ij}(\tau, x_i, y_{ij}(x_i)) d\tau, t \geq T > 0.$$

Parameter

$$\varphi_{ij}^{out}(t) = \frac{w_{ij}^{out}(t)}{d_i^{out}}$$

where $d_i^{out}$ is the output degree of node $n_i$ determines the importance for this node the edge $(n_i, n_j)$.

The flow degrees of nodes are the local dynamic characteristics of the NS.

The functional importance of the edge $(n_i, n_j)$ in the system is determined by the value $V_{ij}(t)$, $i, j = \overline{1, N}$. We will define the functional importance of node in the following way. Denote by $v_k^{out}(t, n_i, n_j)$ the volume of flows generated in node $n_i$ and received at node $n_j$, which passed through the path $p_k(n_i, n_j)$ for the period [$t$-$T$, $t$], $K_{ij}$ is the number of all possible paths that connect nodes $n_i$ and $n_j$, $k = \overline{1, K_{ij}}$, $i, j = \overline{1, N}$. Then

$$V^{out}(t, n_i, n_j) = \sum_{k=1}^{K_{ij}} v_k^{out}(t, n_i, n_j)$$

is the total volume of flows generated in node $n_i$ and directed to accept in node $n_j$ by all possible paths for the period $[t\text{-}T, t]$. Parameter $V^{out}(t, n_i, n_j)$ defines the strength of influence of node $n_i$ on node $n_j$ at the current time $t$, $i, j = \overline{1, N}$. Denote by $J_i^{out} = \{j_{i_1}, ..., j_{i_L}\}$ the set of node numbers that are the final receivers of flows generated in the node $n_i$. Parameter

$$\xi_i^{out}(t) = \sum\nolimits_{j \in J_i^{out}} V^{out}(t, n_i, n_j) / s(\mathbf{V}(t)), \; \xi_i^{out}(t) \in [0,1]$$

determines the strength of influence of node $n_i$ on the system as a whole, $i = \overline{1, N}$. Here $s(\mathbf{M})$ is a sum of elements of the matrix $\mathbf{M} = \{m_{ij}\}_{i,j=1}^{M}$, i.e. $s(\mathbf{M}) = \sum\nolimits_{i=1}^{N} \sum\nolimits_{j=1}^{N} m_{ij}$ and the value $s(\mathbf{V}(t))$ determines the total volume of flows which passed through the network for the period $[t\text{-}T, t]$.

Parameter

$$\psi_{ij}^{out}(t) = V^{out}(t, n_i, n_j) \Big/ \sum\nolimits_{j \in J_i^{out}} V^{out}(t, n_i, n_j)$$

determines for the node-generator $n_i$ the importance of connection with node-receiver of flows.

The power of influence of node $n_i$ on the system is determined by the parameter

$$p_i^{out} = L / N, \; p_i^{out} \in [0,1],$$

where $L$ is the number of elements of the set $J_i^{out}$ which we call the domain of influence of node $n_i$ on the NS, $i = \overline{1, N}$.

So-called botnets are often presented in social online services [24]. By means of these botnets one person can create the illusion of common opinion of many people, massively distribute the disinformation, organize *DdoS*-attacks, and so on. So, in one of the most popular social networks Twitter there are huge networks of fake accounts, the number of nodes of which exceeds 350 thousand [25]. Detection of nodes-generators of such botnets and their blocking allows us to prevent many negative social and economic phenomena. Parameters $\xi_i^{out}(t)$ and $p_i^{out}$, $i = \overline{1, N}$, enable to identify the botnet generators with sufficient precision, since the strength and power of their influence on the NS are usually much higher than average.

Denote by $J^{out}$ the set of all network nodes-generators and introduce parameter

$$V^{out}(t) = \sum\nolimits_{i \in J^{out}} \sum\nolimits_{j \in J_i^{out}} V^{out}(t, n_i, n_j).$$

Obviously, if the rate of flows in the system is high enough and the value $V^{out}(t)$ is less than $s(\mathbf{V}(t))$, this is evidence of the presence of hidden nodes-generators of flows in the system structure; if the value $V^{out}(t)$ is greater than $s(\mathbf{V}(t))$ – an indication of the presence of hidden nodes-receivers of flows.

Determine parameter

$$p^{out} = \mu(J^{out}) / N, \; p^{out} \in [0,1],$$

where $\mu(J^{out})$ is the power (number of elements) of subset $J^{out}$, which determines the specific weight of nodes-generators in the system structure. Obviously, the smaller the value $p^{out}$, the more vulnerable is the NS to destabilization the work of the nodes-generators of flows [16].

Denote by $v_k^{in}(t, n_j, n_i)$ the volume of flows generated in node $n_j$ and received at node $n_i$, which passed through the path $p_k(n_j, n_i)$ for the period [*t-T, t*], $K_{ji}$ is the number of all possible paths that connect nodes $n_j$ and $n_i$, $k = \overline{1, K_{ji}}$, $i, j = \overline{1, N}$. Then

$$V^{in}(t, n_j, n_i) = \sum_{k=1}^{K_{ji}} v_k^{in}(t, n_j, n_i)$$

is the total volume of flows generated in node $n_j$ and directed to accept in node $n_i$ by all possible paths for the period [*t-T, t*]. Parameter $V^{in}(t, n_j, n_i)$ defines the strength of influence of node $n_j$ on node $n_i$ at the current time *t*, $i, j = \overline{1, N}$. Denote by $J_i^{in} = \{j_{i_1}, ..., j_{i_M}\}$ the set of node numbers in which the flows are generated, which are sent for receiving in the node $n_i$. Parameter

$$\xi_i^{in}(t) = \sum_{j \in J_i^{in}} V^{in}(t, n_j, n_i) / s(\mathbf{V}(t)), \; \xi_i^{in}(t) \in [0,1],$$

determines the strength of influence of NS on the node $n_i$, $t \geq T > 0$, $i = \overline{1, N}$.

Parameter

$$\psi_{ij}^{in}(t) = V^{in}(t, n_j, n_i) \Big/ \sum_{j \in J_i^{in}} V^{in}(t, n_j, n_i)$$

determines for the node-receiver $n_i$ the importance of connection with node-generator of flows.

The power of influence of the system on the node $n_i$ is determined by the parameter

$$p_i^{in} = M / N, \; p_i^{in} \in [0,1],$$

where *M* is the number of elements of the set $J_i^{in}$ which we call the domain of influence of NS on the node $n_i$, $i = \overline{1, N}$. In social networks, parameters $\xi_i^{in}(t)$ and $p_i^{in}$, $i = \overline{1, N}$, allow us to identify users whose judgments pose the greatest attention of the Internet community, since the response to them (the strength and power of influence from the NS) is significantly higher than average.

Denote by $J^{in}$ the set of all network nodes-receivers and introduce parameter

$$V^{in}(t) = \sum_{i \in J^{in}} \sum_{j \in J_i^{in}} V^{in}(t, n_j, n_i).$$

Obviously, if the rate of flows in the system is high enough and the value $V^{in}(t)$ is less than $s(\mathbf{V}(t))$, this is evidence of the presence of hidden nodes-receivers of flows in the system structure; if the value $V^{in}(t)$ is greater than $s(\mathbf{V}(t))$ – an indication of the presence of hidden nodes-generators of flows.

Determine parameter

$$p^{in} = \mu(J^{in}) / N, \; p^{in} \in [0,1],$$

where $\mu(J^{in})$ is the power of subset $J^{in}$, which determines the specific weight of nodes-receivers in the system structure. Obviously, the smaller the value $p^{in}$, the more vulnerable is the NS to destabilization the work of the nodes-receivers of flows.

Parameters of the strength and power of influence are global dynamic characteristics of node in the NS. Attacks on the nodes with large values of these parameters can significantly destabilize the whole system or a large part of it. Special attention in TCN is given to the issue of network stability, as its ability to resist targeted external influences (hacker or terrorist attacks, etc.) [26]. The other side of system stability consist in its sensitivity to small changes in the structure or operation process. Such changes can be caused by both internal and external factors, and can lead to the no less consequences than targeted attacks. In this case, the stability of structure is determined by the sensitivity to small changes in the set of its nodes and edges. The structure is unstable when such changes can lead to loss of certain network properties, such as connectivity. The stability of NS operation process is determined by its sensitivity to small changes in the volume of flows movement. For example, the systems operation may become unstable in the conditions of critical loading of part of its edges (the corresponding elements of matrices $\mathbf{u}(t)$ or $\mathbf{U}(t)$ are close to 1) or some the most important nodes in terms of strength and power of influence. Many systems are sensitive to small violations of established schedule of flows motion. Obviously, the stability of process is associated with the stability of NS structure. If small changes (blocking some network nodes and edges) lead to loss of connectivity, this directly affects on the systems operation. If the load of certain elements of structure by flows is critical (close to their bandwidth), it also creates a threat of blocking these elements.

Node $n_i$, for which $\xi_i^{in}(t) = \xi_i^{out}(t) = 0$ and $W_i^{in}(t) = W_i^{out}(t) = W_i^{tr}(t) \neq 0$, $t \geq T > 0$, $i = \overline{1, N}$, will be called a transit node. The importance of transit node in the system is determined by the volume of flows that pass through it. Extraction from structure the transit nodes is one way to reduce the dimensionality of system model. It should be borne in mind that destabilization of important transit node operation with large value $W_i^{tr}(t)$ and high betweenness centrality can destabilize the whole system or large part of it [27].

The preferential influence $\psi_i(t)$ of node $n_i$ for non-transit NS nodes we will determined by the ratio

$$\psi_i(t) = (\xi_i^{in}(t) - \xi_i^{out}(t))/(\xi_i^{in}(t) + \xi_i^{out}(t)), \ \psi_i \in [-1, 1].$$

If the value of parameter $\psi_i(t)$ is close to –1, then the preferential influence is from the node $n_i$ on NS. If the value of parameter $\psi_i(t)$ is close to 1, then the preferential influence is from NS on the node $n_i$. In case $\psi_i(t) \approx 0$, $i = \overline{1, N}$, the influence is uniform on each side. The network structure (Fig. 3a) is usually much simpler than the structure of flows in it (Fig. 3b). Parameter of preferential influence allows us to determine the predominant direction of flows in the system (Fig. 3c). Thus, passenger traffic in a country or a large city is characterized by the value of $\psi_i(t) \approx 0$, $i = \overline{1, N}$. At the same time, migration processes (refugee movement, urbanization, etc.) are characterized by a pronounced uneven distribution of the values of preferential influence.

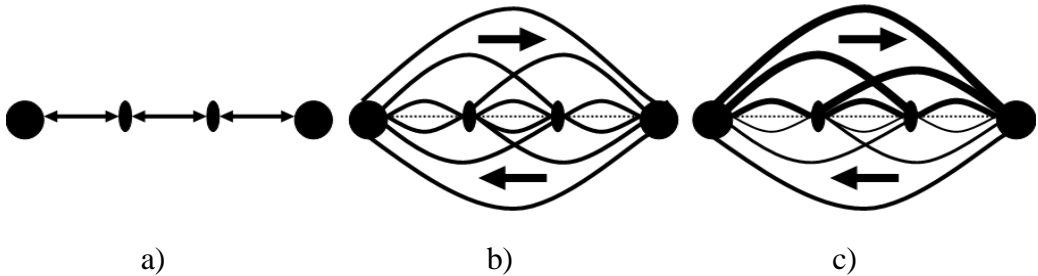

a)  b)  c)

Fig. 3. Fragvents: a) network structure; b) structure of flows in network; c) volumes of flows motion in network.

IV. FLOW CORES OF NETWORK SYSTEMS

We looked above two methods of reducing the dimensionality of NS models, which consisted in removing from the system structure the fictitious nodes and edges, and transit nodes. Another way to simplify the network model is to introduce the notion of its *k*-core, that is, the largest subnet of the source CN, all nodes of which have degree not less than *k*, and the extraction from the network structure of nodes with degree less than *k* [28]. Using the flow characteristics of NS allows us to introduce the concept of flow $\lambda$-core of network system, as the largest subnet of source network, for which all elements of the integral flow adjacency matrix have values not less than $\lambda$, $\lambda \in [0, 1]$. Fig. 4a reflects the structure of railway transport system of the western region of Ukraine. The thickness of lines in this figure is proportional to the weight of edges – the volumes of flows passing through them. In total, this network contains 354 nodes. Fig. 4b displays the 4-core of this network, which contains 12 nodes and 28 edges. Fig. 4c displays the flow 0.7-core of this system, which contains 4 nodes and 12 edges.

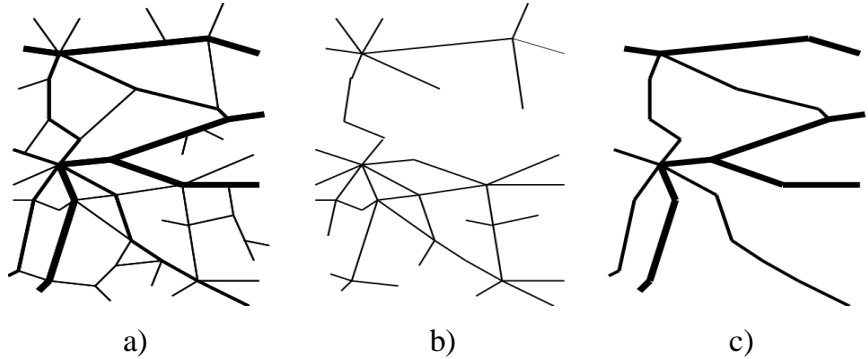

a)          b)          c)

Fig. 4. Fragments: a) weighted network; c) 4-core of CN; c) flow 0,7-core of NS.

The dimensionality of large NS models can be reduced by investigation not source network, but only its $\lambda$-core. Moreover, the greater the specific weight of $\lambda$-core in the system, the more adequate the result of study. Introduce the IFAM of $\lambda$-core by means of ratio

$$\mathbf{V}^\lambda(t) = \{V_{ij}^\lambda(t)\}_{i,j=1}^N, V_{ij}^\lambda(t) = \begin{cases} V_{ij}(t), \text{ if } V_{ij}(t) \geq \lambda, \\ 0, \text{ if } V_{ij}(t) < \lambda. \end{cases}$$

We will use parameter $\sigma_\lambda(t)$ to determine the specific weight of $\lambda$-core. This parameter is equal to the ratio of volumes of flows passing by the $\lambda$-core to the volume of flows that pass through the network as a whole during the period $[0, T]$:

$$\sigma_\lambda(t) = s(\mathbf{V}^\lambda(t))\big/s(\mathbf{V}(t)).$$

Since the main goal of the most network systems is to provide the flows motion, parameter $\sigma_\lambda(t)$ quantifies how the $\lambda$-core provides the implementation of this goal. If instead of model the source system we are investigating the model of its $\lambda$-core, then the value $\sigma_\lambda(t)$ can be interpreted as a dynamic measure of adequacy of this model. Selection and study of the flow NS's core compared to the whole network research can reduce the dimensionality of model by rejecting the least functionally important nodes and edges. As a result, research of network is reduced to the analysis or modeling of the most priority its substructures. So, the spread of epidemics usually occurs on the ways of intensive movement of large masses of people, and the spread of computer viruses - on the paths of intense information traffic. The flow cores of NS with large values of $\lambda$ determine the most likely paths of deploying such processes.

Consider some source network. Determine its $\lambda$-core and *k*-core of this $\lambda$-core. As a result, we obtain a subnet of the source CN, which we will call $\lambda(k)$-core of NS. The procedure for defining the $\lambda(k)$-core of NS consisted in the primary exclusion of edges for which the corresponding elements of IFAM are smaller than $\lambda$, reduces the probability of obtaining unconnected structures.

Obviously, the $\lambda$-cores with lagre values $\lambda$ allow us to determine the most important paths in the system structure.

The cardiovascular system of human body includes both the main and peripheral veins and arteries. A rupture of one of the major vessels can lead to quick fatal consequences. Overlap the main highways of large city can lead to collapse of its transport system. Accidents on trunk power lines during natural disasters often lead to the disconnection of electricity in entire regions of the country. However, this does not mean that connections with the small value $\lambda$ can be completely ignored. In human society and physical world the great importance has so-called weak interactions due to which the "small world" networks may exist [29]. The number of small settlements is prevalent in any country. These settlements need to be provided with products, transportation, financial, medical, educational and other services, despite the small volume of flows. If the length of large vessels (veins, venules, arteries and arterioles) of the human body reaches several kilometers, then the length of its capillary network is close to 100 thousand km (Fig. 5a) [30]. However, the capillary network supplies the human body cells with all essential substances for life activity and removes the results of this activity. That is, the main function of cardiovascular system is realized precisely by the capillary network, and large vessels act as flow conveyors. Similar considerations are true for many other types of NS.

The above examples lead us to another way of reducing the size of NS models consisted in encapsulation of some subnets of the source CN. For the cardiovascular system this means reducing the dimensionality its model in a thousand times (Fig. 5b). The encapsulation is expedient to use in many real systems (communities of social nature [26], infrastructure of settlements in the motor transport systems of region or country etc.).

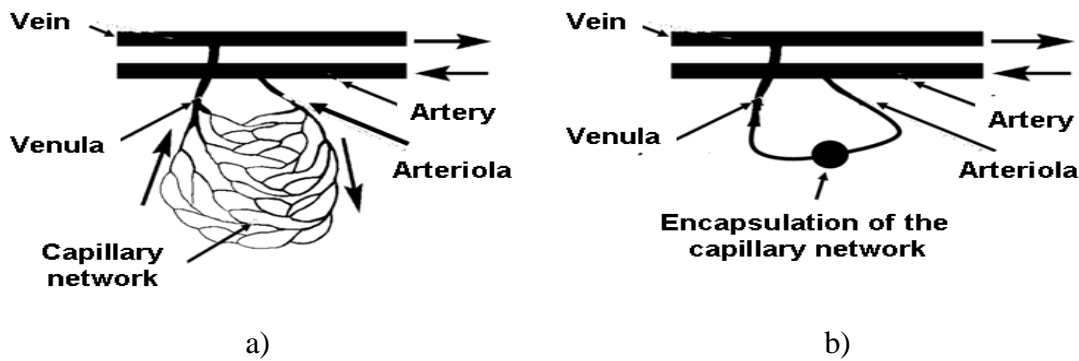

Fig. 5. Encapsulation of capillary subnetworks of the cardiovascular system.

List the main features of subnet that can be encapsulated:

1) it is a connected component of complement to separate $\lambda$-core of source network;
2) each element of this subnet has a negligible influence on the system and vice versa, but the influence of encapsulated domain is commensurate with the influence of other components of $\lambda$-core;
3) the number of connections of the encapsulated domain does not exceed the maximum degree of nodes in source network.

Obviously, the encapsulation of subnets of source network can significantly enrich the model of $\lambda$-core and increase its adequacy.

## V. CONCLUSIONS

In this article, the structural and functional approaches of systems research are reviewed. The analysis is provided for qualitative and quantitative aspects of systems and system models complexity. It is determined that approaching functional (qualitative) complexity we have to set aside its structural (quantitative) complexity. It is shown that functional approach allows us to study peculiarities of both structure and system development at all stages of their life cycle, not just on the stage of growth. In order to study the process of network systems operation the flow adjacency matrices of different types were introduced. It was also analyzed, how these matrices help to study and forecast the peculiarities of this process, evaluate its efficiency and prevent existing and potential threats. Local and global dynamic characteristics of the network systems elements were determined, such as input and output node flow degrees, functional importance of CN's nodes and edges, strength and power of separate node's influence on system and system's influence on separate node. These characteristics allow us to determine the predominant direction of flows within the system, identify nodes that generate and receive flows, and transit nodes, study activity, passivity, critical load and stability of separate NS components and NS in general, enable to define the actual structure of the system by searching fictitious and hidden nodes and edges, thus decreasing the dimensionality of model and increasing its adequacy. The notion of flow core of the network system was introduced and its specific weight in the NS was determined as the quantitative measure of the corresponding model adequacy. The methods were analyzed for encapsulation of separate components of network structures which allow us to substantially reduce system dimensionality.

REFERENCES

[1] N. Boccara, Modeling Complex Systems. Springer Science & Business Media, 2010.
[2] J. Smith and C. Jenks, Qualitative Complexity: Ecology, Cognitive Processes and the Re-Emergence of Structures in Post-Humanist Social Theory. Routledge, 2006.
[3] W. H. Zurek, Complexity, Entropy And The Physics Of Information. CRC Press, 2018.
[4] C. Spickermann, Entropies of condensed phases and complex systems. Springer, Berlin, 2011.
[5] M. J. Eppler and J. Mengis, "The Concept of Information Overload: A Review of Literature from Organization Science, Accounting, Marketing, MIS, and Related Disciplines", The Information Society, vol. 20(5), pp. 325-344, May 2010.
[6] A.-L. Barabási and J. Frangos, Linked: the new science of networks. Basic Books, New York, 2002.
[7] D. Polishchuk, O. Polishchuk and M. Yadzhak, "Solution of some problems of evaluation of the complex systems", 15$^{th}$ Int. conf. on automatic control, 23–26 September 2008, Odesa, pp. 968–976.
[8] D. Polishchuk, O. Polishchuk and M. Yadzhak, "Complex Evaluation of Hierarchically-Network Systems", Automatic Control and Information Sciences, vol. 2(2), pp. 32–44, May 2014.
[9] Y.-Y. Liu, J. J. Slotine and A. L. Barabási, "Observability of complex systems", Proc. of the National Academy of Sciences, vol. 110(7), pp. 2460-2465, July 2013.
[10] S. Boccaletti, V. Latora, Y. Moreno, M. Chavez, D. U. Hwang, "Complex networks: Structure and dynamics", Physics reports, vol. 424(4), pp. 175-308, Apr 2006.
[11] Yu. Holovatch, O. Olemskoi, C. Von Ferber, T. Holovatch, O. Mryglod, I. Olemskoi, V. Palchykov, "Complex networks", Journal of Physical Studies, vol. 10 (4), pp. 247-289, Dec 2006.
[12] S. N. Dorogovtsev and J. F. F. Mendes, Evolution of Networks: From Biological Nets to the Internet and WWW. Oxford University Press, Oxford, 2013.
[13] G. Caldarelli and A. Vespignani, Large Scale Structure and Dynamics of Complex Networks: From Information Technology to Finance and Natural Science. World Scientific, New York, 2007.
[14] R. G. Coyle, System Dynamics Modelling: A practical approach. CRC Press, 1996.
[15] D. Polishchuk, O. Polishchuk, and M. Yadzhak, "Complex deterministic evaluation of hierarchically-network systems: I. Methods description", System Research and Information Technologies, vol. 1, pp. 21-31, Mar. 2015.
[16] O. Polishchuk and M. Yadzhak, "Network structures and systems: I. Flow characteristics of complex networks", System research and informational technologies, vol. 2, pp. 42-54, July 2018.
[17] A. Barrat, M. Barthélemy and A. Vespignani, "The Architecture of Complex Weighted Networks: Measurements and Models", In Large Scale Structure and Dynamics of Complex Networks. World Scientific, pp. 67-92, 2007.

[18] M. E. J. Newman, "Analysis of weighted networks", Phys. Rev. E, vol. 70, 056131, Aug 2004.

[19] R. Albert and A.-L. Barabasi, Statistical mechanics of complex networks, Review of Modern Physics, vol. 74 (1), Jan 2002.

[20] O. Polishchuk, "Flows characteristics and cores of complex network and multiplex type systems", arXiv preprint, arXiv:1702.02730 [physics.soc-ph], 9 Feb 2017, 22 p.

[21] O. Polishchuk, M. Tyutyunnyk, and M. Yadzhak, "Quality evaluation of complex systems function on the base of parallel calculations, Information Extraction and Processing, vol. 26 (108), pp. 121-126, Dec 2007.

[22] C. Prell, Social Network Analysis: History, Theory and Methodology. SAGE, New York, 2012.

[23] G. Price and C. Sherman, The Invisible Web: Uncovering Information Sources Search Engines Can't See. CyberAge Books, New York, 2001.

[24] Q. Cao, M. Sirivianos, X. Yang, and T. Pregueiro, "Aiding the Detection of Fake Accounts in Large Scale Social Online Services", 9th USENIX Symposium on Networked Systems Design and Implementation, San Jose, CA, 2012, pp. 197-210.

[25] N. Abokhodair, D. Yoo, and D. W. McDonald, "Dissecting a Social Botnet: Growth, Content and Influence in Twitter", 18th ACM Conference on Computer Supported Cooperative Work & Social Computing, Vancouver, BC, Canada, 2015, pp. 839-851.

[26] M. Newman, Networks. An Introduction. Oxford University Press, 2010.

[27] O. Polishchuk and D. Polishchuk, "Monitoring of flow in transport networks with partially ordered motion", XXIII Conf. Carpenko physics and mechanics institute, NASU, 2013, pp. 326-329.

[28] S. N. Dorogovtsev, A. V. Goltsev, and J. F. F. Mendes, "k-core organization of complex networks", Physical review letters, vol. 96(4), 040601, Apr 2006.

[29] D. J. Watts and S. H. Strogatz, "Collective dynamics of 'small-world' networks", Nature, vol. 393, pp. 440–442, Feb 1998.

[30] M. K. Pugsleya and R. Tabrizchib, "The vascular system: An overview of structure and function", Journal of Pharmacological and Toxicological Methods, vol. 44(2), pp. 333-340, Feb 2000.